\documentstyle[amssymb,aps,prl,floats,epsfig,twocolumn]{revtex}

\begin{document}
\draft

\wideabs{
\title{Alkali specific effects in superconducting fullerides: the observation of a
high temperature insulating phase in Na$_{2}$CsC$_{60}$}
\author{Natasa \v{C}egar\cite{natasa}, Ferenc Simon\cite{feri}, Slaven Garaj,
L\'{a}szl\'{o} Forr\'{o}}
\address{Laboratoire de Physique des Solides Semicristallins, IGA Department de\\
Physique, Ecole Polytechnique Federal de Lausanne, 1015 Lausanne, Switzerland}
\author{Barbara Ruzicka, Leonardo Degiorgi}
\address{Laboratorium f\"{u}r Festk\"{o}rperphysik, ETH-\\
Z\"{u}rich CH-8093 Z\"{u}rich, Switzerland}
\author{Veronique Brouet}
\address{Laboratoire de Physique des Solides, 91405 Orsay, France}
\author{L\'{a}szl\'{o} Mih\'{a}ly}
\address{SUNY Stony Brook, Department of Physics, Stony Brook, NY 11794 USA}
\date{\today}
\maketitle

\begin{abstract}
Electron Spin Resonance and optical reflectivity measurements demonstrate a
metal-insulator transition in Na$_{2}$CsC$_{60}$ as the system passes from
the low temperature simple cubic to the high temperature {\it fcc} structure
above 300 K. The non-conducting electronic state is especially unexpected in
view of the metallic character of other, apparently isostructural
fullerides, like K$_{3}$C$_{60}$. The occurence of this phase in Na$_{2}$CsC$_{60}$ suggests that alkali specific effects can not be neglected in the
description of the electronic properties of alkali doped fullerides. We
discuss the origin of the insulating state and the relevance of our results
for the anomaly observed in the magnitude of the superconducting transition
temperature of Na$_{2}$AC$_{60}$ fullerides.
\end{abstract}

}

Since the original discovery of superconductivity in A$_{3}$C$_{60}$ (A=K,
Rb)\cite{haddon91}, a great progress has been made in understanding this
phenomenon in fullerides, however the detailed description of the normal
state properties is still missing. There is no consensus about the role of
Coulomb correlations, about the importance of polaronic effects and
Jahn-Teller distortion, and about the influence of the dopants on the
electronic properties. Expanding the lattice by intercalation or by
temperature causes a metal-insulator (M-I) transition, but it is not clear
whether this is a simple Mott-Hubbard transition or lattice effects are also
involved. It has become more urgent to answer these questions with the
synthesis of fullerides with lighter alkali atoms like Li or Na, e.g. Li$%
_{x} $CsC$_{60}$, Na$_{2}$CsC$_{60}$ etc. In these systems, even the simple
relationship between the superconducting transition temperature and the
lattice spacing which worked very well for A$_{3}$C$_{60}$ seems to fail\cite
{ramirez-fleming}\cite{sparn}\cite{yildirimssc95}. According to the widely
accepted description of superconductivity in triply charged cubic
fullerides, the alkali atoms act only to expand the lattice and there is no
alkali specific effect on the electronic properties of the compounds. The
differences between various systems are attributed solely to the lattice
constant ($a$) dependence of the density of states at the Fermi level, N(E$%
_{F}$). \ In simple cubic ({\it sc}, space group Pa$\overline{3}$) Na$_{2}$AC%
$_{60}$ (A=K, Rb, Cs) compounds, however, T$_{c}$ increases much steeper
with $a$ than in the other A$_{3}$C$_{60}$ fullerides with the Fm$\overline{3%
}$m structure\cite{gunnarssonrmp97}. This observation may be explained if a
stronger N(E$_{F}$) vs. $a$ dependence is postulated in Na$_{2}$AC$_{60}$
compounds\cite{yildirimssc95}. Yet band calculations give similar N(E$_{F}$)
vs. $a$ dependence for the two structures\cite{gunnprb95}, and at least two
experimental results contradict the stronger variation of N(E$_{F}$) on $a$:
i.) NMR measurements found that N(E$_{F}$) is not reduced sufficiently to
explain for the reduction of T$_{c}$ in Na$_{2}$RbC$_{60}$ and Na$_{2}$KC$%
_{60}$\cite{maniwaprb95}\cite{saitocomment}; ii.) The variation of T$_{c}$
under pressure in Na$_{2}$CsC$_{60}$\cite{mizukiprb94} and Na$_{2}$Rb$_{0.5}$%
Cs$_{0.5}$C$_{60}$\cite{brownprb99} suggests that N(E$_{F}$) depends on $a$
similarly to the A$_{3}$C$_{60}$ compounds.

The T$_{c}$($a$) curve thus remains unsatisfactorily explained and motivates
a detailed comparison of Na$_{2}$AC$_{60}$ and other A$_{3}$C$_{60}$
fulleride superconductors. The Na$_{2}$AC$_{60}$ compounds undergo a {\it sc}%
-{\it fcc} structural transition around room temperature (T$_{s}=299$ K for
Na$_{2}$CsC$_{60}$)\cite{tanigakiprb94} and become isostructural to other A$%
_{3}$C$_{60}$ fullerides at high temperatures, making a direct comparison of
the electronic properties possible. For Na$_{2}$CsC$_{60}$ a jump in the
cubic lattice constant accompanies the structural transition\cite
{kosmasscience94}, but the lattice ({\it a}($T$=425 K)=14.1819 \AA \cite
{kosmasscience94}) is still contracted relative to other A$_{3}$C$_{60}$
compounds ({\it a}($T$= 300 K)=14.240 \AA\ for K$_{3}$C$_{60}$\cite
{stephensnature91} that has the smallest lattice constant among the A$_{3}$C$%
_{60}$ fullerides). The overlap between adjacent C$_{60}$ balls must be
larger in Na$_{2}$AC$_{60}$ than in A$_{3}$C$_{60}$ compounds. Thus, Na$_{2}$%
AC$_{60}$ systems are less prone to correlation effects and are expected to
be also metals in the {\it fcc} phase like the A$_{3}$C$_{60}$ fullerides 
\cite{palstraprb94}. In sharp contrast to this expectation, here we report
that Na$_{2}$CsC$_{60}$ {\em is not a metal} in the high temperature {\it fcc%
} phase. Electron Spin Resonance (ESR) on Na$_{2}$CsC$_{60}$ (as compared to
K$_{3}$C$_{60}$) and the infrared reflectivity (IR) studies demonstrate that
the high temperature {\it fcc} phase of Na$_{2}$CsC$_{60}$ is a gapped
insulator. The difference between Na$_{2}$CsC$_{60}$ and K$_{3}$C$_{60}$
suggests that alkali effects are important and provide further input for the
solution of the long lasting puzzle related to the T$_{c}$ vs. $a$ anomaly
in Na$_{2}$AC$_{60}$ fullerides.

Several Na$_{2}$CsC$_{60}$ samples were prepared by conventional solid-state
reaction method. X-ray diffraction showed them to be single phase. We
studied powder samples by ESR, and powder and pressed pellet samples by IR.
The powder samples are superconductors with T$_{c}$=11.7 K. The pellets were
made at 10 kbar pressure at room temperature. The shielding fraction of 22\%
at 4.2 K of the powder sample diminishes to less than 0.1 \% in the pressed
pellet\cite{gyuri}. The pressed pellets are non-superconducting polymers\cite
{margadonnaJSSC} that are isostructural to the polymeric phases of Na$_{2}$KC%
$_{60}$\cite{prassidesphysc97} and Na$_{2}$RbC$_{60}$\cite{bendeleprl}. For
the powder sample (sealed in quartz tubes with a low pressure He exchange
gas) X-band ESR($\sim $9 GHz) experiments were performed in the 5-800 K
temperature range. High flux of flowing exchange gas was used for
thermalization of the sample-holder and variation of cavity parameters were
carefully monitored. No sample degradation was observed up to 800 K. The ESR
signal of previously studied K$_{3}$C$_{60}$ powder samples \cite{norbi} was
also recorded, and it served as a reference for comparison. IR measurements
below 300 K were performed on Na$_{2}$CsC$_{60}$ powder samples in a sealed
sample holder with wedged diamond window. The highest $T$ (300 K)\ of the
IR\ apparatus prevents a study of the other two Na$_{2}$AC$_{60}$ (A=K or
Rb) compounds ($T_{s}$=321 K for Na$_{2}$KC$_{60}$\cite{simonunpub} and 313
K for Na$_{2}$RbC$_{60}$\cite{tanigakiprb94}) in the {\it fcc} phase. Grains
of the powder sample do not form a homogeneous and compact specimen thus
scattering of the incident light and transparency of the probe limited the
spectral range to 200 cm$^{-1}-$ 6000 cm$^{-1}$. In case of the pressed
pellet samples, the reflectivity, $R$($\omega $), was measured from 30 cm$%
^{-1}$ up to 5$\times $10$^{3}$ cm$^{-1}$\cite{bommeli}. IR reflectivity
data was calibrated by an Al mirror reference. The optical conductivity $%
\sigma $($\omega $) = $\sigma _{1}$($\omega $) + i$\sigma _{2}$($\omega $)
was obtained from Kramers-K\"{o}nig (KK) transformations of the measured
reflectivity. Standard extrapolations were used above our highest frequency
limit, while below the lowest measured frequency we performed Hagen-Rubens
extrapolation for the metallic phase and the extrapolation to a constant
value of $R$($\omega $) in the case of an insulating behavior.

In Fig. 1. we show the spin-susceptibility ($\chi \left( T\right) $) and ESR
linewidth ($\Delta H\left( T\right) $) of Na$_{2}$CsC$_{60}$. The
susceptibility and linewidth of K$_{3}$C$_{60}$, normalized to the 300 K
values, is also shown for comparison. The ESR signals of our Na$_{2}$CsC$%
_{60}$ and K$_{3}$C$_{60}$ samples are identical to those reported in Ref. 
\cite{petitprb96} and Ref.\cite{petitprb98} (studied between 25-300 K). We
discuss three separate $T$ regions: i.) 5-12 K where the system is a
superconductor; ii.) 12-300 K where the system shows metallic behavior;
iii.) 300-800 K, the insulating {\it fcc} phase.

Superconductivity below $T_{c}\simeq $ 12 K is marked by a drop of $\chi
\left( T\right) $ as a result of the microwave field exclusion from the
sample. No vortex noise was observed down to 5 K,\ indicating that the
sample was in the vortex-liquid state. The ESR line broadens on lowering $T$
below $T_{c}$ due to the development of magnetic field inhomogeneities in
the sample. In the 12-300 K region conduction electron spin resonance is
observed. The increase of $\chi \left( T\right) $ on increasing $T$ is not
typical for a conventional metal but is a common feature of alkali fulleride
metals in the normal state\cite{petitprb98}. The origin of this variation is
still unclear, although attempts were made to assign it to the variation of
N(E$_{F}$) due to the varying $a$\cite{petitprb98}. The linewidth, $\Delta
H\left( T\right) $, follows a $T$ dependence that is common for a metal when
lattice vibrations dominate the relaxation of conducting electrons\cite
{petitprb96}. The {\it sc}-{\it fcc} structural transition at $T_{s}=299$ K 
\cite{tanigakiprb94} appears as a minimum in $\Delta H\left( T\right) $ and
a maximum in $\chi \left( T\right) $ around $T\simeq $ 300 K. The minimum in
\ $\Delta H\left( T\right) $ \ occurs when the frequency of the structural
fluctuations is equal to the Larmor frequency. The character of the
electronic properties clearly changes as $\Delta H\left( T\right) $ no
longer follows a linear behavior but increases slightly and saturates at
higher $T$. The variation of $\chi \left( T\right) $ above $T_{s}$ follows a
quantitatively different behavior from that of K$_{3}$C$_{60}$: $\chi \left(
T\right) $ of the K$_{3}$C$_{60}$ metal changes little, whereas $\chi \left(
T\right) $ of Na$_{2}$CsC$_{60}$ drops by a factor $\sim $2 between 300 and
800 K. This drop is reproducible and no hysteresis is observed. Such a large
variation is difficult to reconcile with any models based on a metallic band
picture. Thus our ESR data is suggestive of the presence of localized
paramagnetic moments even though $\chi \left( T\right) $ data alone can not
provide an unambiguous identification of the electronic state. The
properties of the ESR signal of the other two members, Na$_{2}$KC$_{60}$ and
Na$_{2}$RbC$_{60}$, of the Na$_{2}$AC$_{60}$ system were identical (i.e.
substantially decreasing $\chi \left( T\right) $ and saturating $\Delta
H\left( T\right) $ above $T_{s}$\cite{simonunpub}\cite{ownkirchberg}) to
that of Na$_{2}$CsC$_{60}$.

The nature of the {\it fcc} phase of Na$_{2}$CsC$_{60}$ is clarified by the
reflectivity spectra, $R$($\omega $), and the corresponding real part, $%
\sigma _{1}$($\omega $), of the optical conductivity for the powder sample
at some relevant temperatures (Fig. 2). The insulator-metal transition is
clearly evidenced by the optical reflectivity. Indeed, $R$($\omega $) at 300
K presents a flat insulating-like spectrum without any clear plasma edge
onset and suggesting an extrapolation to a constant value for $\omega
\rightarrow $ 0. Below 220 K we see, on the other hand, the onset of the
plasma edge feature at frequencies around 50 meV, that is a typical optical
fingerprint for a metallic behavior. Such an onset becomes more and more
pronounced as approaching 20 K. It appears that, even though the {\it sc}-%
{\it fcc} transition takes place at $T_{s}=299$ K, its optical manifestation
is clearly seen only below about 220 K. An {\it ad hoc} Lorentz-Drude fit 
\cite{Wooten}, which reproduces the plasma edge onset, allows us to
extrapolate $R$($\omega $) in a metallic-like fashion for $\omega
\rightarrow $ 0 at 200 and 20 K(See Fig. 2.). The extrapolated dc
conductivity is of about 100 $\Omega ^{-1}$cm$^{-1}$ ($\sim $10 times
smaller than in K$_{3}$C$_{60}$\cite{degiorgi})$.$ At 300 K, $\sigma _{1}$($%
\omega $) (Fig. 2b.) increases with increasing frequency, showing a first
bump at 70 meV and the onset of an absorption peaked around 0.3 eV. At
temperatures below 220 K, Drude weight appears at low frequencies (i.e.,
below 40 meV), while the absorptions decrease in intensity. It must be
noticed that the apparent non-conservation of the spectral weight is mainly
the consequence of the high frequency extrapolation of $R$($\omega $) for
the purpose of the KK transformation. Since the measurements at different
temperatures do not merge together (see Fig. 2a.), a different extrapolation
was employed at 20, 220 and 300 K. The standard way is to extend the
measured $R$($\omega $) beyond 0.6 eV with a suitable combination of Lorentz
harmonic oscillators\cite{Wooten} so that $R$($\omega $) $\rightarrow $ 0
for $\omega \rightarrow $ 0. Depending from the mode strength of these
oscillators, the spectral weight encountered by the broad feature at 0.3 eV
and the peak maximum might change and shift, respectively.

The measurements of the polymerized phase in the pressed pellet samples, as
shown in the inset of Fig. 2b., prove the sensitivity of the optical
measurements with respect to different phases. The polymerized phase of Na$%
_{2}$CsC$_{60}$ is a metal at all $T$ as $\sigma _{1}$($\omega $) is
characterized by a low frequency Drude term\cite{leotocome}. The Drude
weight and correspondingly the dc limit of $\sigma _{1}$($\omega $)
decreases with decreasing $T$. This suggests a disordered-metal like
scenario for the Na$_{2}$CsC$_{60}$ polymer, as in non-oriented doped
polymers\cite{bommeli}. We recall, that although the polymerized phase is
always metallic, it is not superconducting\cite{gyuri}.

The temperature dependent ESR susceptibility and IR studies establish that Na%
$_{2}$CsC$_{60}$ is an insulator in its {\it fcc} phase. This is a striking
result in comparison with the isostructural and larger lattice constant K$%
_{3}$C$_{60}$ metal. This result, in our opinion, may be attributed to the
small size of the Na$^{+}$ ion and consequently to its mobility that
modifies the electronic properties. We propose two alternative models that
both give the phenomenological description of the experimental observations
and are based on the mobility of Na$^{+}$. One possible way of interpreting
the experimental results is in the framework of the Mott-Jahn-Teller
insulating state proposed for A$_{3}$C$_{60}$ by Tosatti and co-workers\cite
{tosatticondmat}. This scenario predicts an insulating magnetic state with
low spin, S=1/2 spin/C$_{60}$ in A$_{3}$C$_{60}$ if the ratio of the on-site
Coulomb interaction (U) and bandwidth (W), (U/W), is larger than a critical
value, (U/W)$_{cr}$. The magnetic susceptibility of this phase follows a $%
\chi (T)=C/(T+T_{N})$ Curie-Weiss temperature dependence above $T_{N}$,
where C is the Curie constant and $T_{N}$ is the N\'{e}el temperature of the
order of W$^{2}/$U. For the broad range of values of W and U that are
available in the literature\cite{gunnarssonrmp97}, $T_{N}$ would be a few
hundred K. In Fig. 1a. we show (solid curve) the simulated high temperature
susceptibility of \ {\it fcc }Na$_{2}$CsC$_{60}$ with Curie constant
corresponding to 1 $\mu _{B}/$C$_{60}$\cite{momentremark} and $T_{N}=200$ K.
We lack independent measurements to determine these parameters separately.
Nevertheless, the value of the magnetic moment only weakly depends on the
parameter $T_{N}$. Thus, we may conclude that a low spin state is realized
in agreement with the theorethical prediction\cite{tosatticondmat}. In the 
{\it fcc}\ phase of Na$_{2}$CsC$_{60}$ the ratio (U/W) must be larger than
in K$_{3}$C$_{60}$ (the latter one being a metal). In view of the smaller
lattice spacing, the bandwidth must be larger. There is no reason to believe
that the U is considerably different in Na$_{2}$CsC$_{60}$ \ than in any
other A$_{3}$C$_{60}$ \ compound. We propose that the formation of mobile
polaron-like Na$^{+}$ entities might reduce the bandwidth thus increasing
U/W over the critical value.

Our alternative scenario is also based on the specific role of the mobile Na$%
^{+}$ ions. The Jahn-Teller effect is known\cite{victoroff} to help the
(static) charge dissociation of the molecular C$_{60}^{3-}$ \ into C$%
_{60}^{4-}$ and C$_{60}^{2-}$, causing a M-I transition, yet in the {\it sc}
phase the the material is a metal. This case we may explain our data due to
an enhancement in this effect by the mobility of Na$^{+}$ ions that promotes
the dissociation of C$_{60}^{3-}$ into C$_{60}^{4-}$ and C$_{60}^{2-}$ for
molecules close or further away from Na$^{+}$, respectively. The Na$^{+}$
are less mobile and are probably locked in the ordered structure to higher
electron density positions in the {\it sc} phase. Thus, the dissociation
effect is less effective than in the {\it fcc} phase. In the {\it fcc} phase
the large thermal motion of the Na$^{+}$, and the disappearance of favored
high electron density positions, enhance the dissociation. The quadruply and
doubly charged molecules are diamagnetic and the excited states (that lie in
the range of 250 meV\cite{kerkoud}) are little accessible in the temperature
ranges of our experiment. The dropping $\chi (T)$ with increasing $T$ is
thus not inherent in this model but the enhanced mobility of Na$^{+}$ at
higher $T$ further reduces $\chi (T)$. This ''alkali polaron enhanced
Jahn-Teller effect'' model explains the two observations of the ESR signal:\
the continuously dropping $\chi (T)$ and saturating $\Delta H\left( T\right) 
$. The charge gapped behavior that we observe in the IR\ experiment is
explained even if a partial dissociation happens as the dissociation is
dynamic and reduces electron hopping, which hinders the conducting electron
percolation in the system. Note that this model, although with less mobile
anions of K$^{+}$ (and in consequence with the dissociation effect in lesser
extent), can account for the ''concave'' shape of the temperature dependence
of the spin susceptibility of K$_{3}$C$_{60}$, as well.

In conclusion, our data reveal a sequence of insulating, metal,
superconducting phase transitions in Na$_{2}$CsC$_{60}$ on lowering
temperature. This is the first observation of an insulating state developing
at higher temperature among the triply charged cubic alkali doped
fullerides. We present two interpretation for the insulating state: a
Mott-Jahn-Teller {\it or} dynamic charge dissociation promoted by the light
Na$^{+}$ ions. Our result can also give a hint about the reasons for the
anomalous T$_{c}$ vs. {\it a} slope in {\it sc} Na$_{2}$AC$_{60}$
superconductors. Since the spin-susceptibility for the whole series of A=Cs,
Rb, K is in the (3.5$\pm 0.5)$*10$^{-4}$ emu/mole range\ (at 100 K)\cite
{simonunpub}, N(E$_{F}$) is not varying strongly enough with $a$, thus the
strong suppression of T$_{c}$ can not be a density of states effect. It is
likely that the lighter alkali ions can respond more easily to the higher
electronic density regions in the {\it sc} lattice, modifying the
electronic/vibrational properties of the C$_{60}^{3-}$ molecule to such an
extent, that superconductivity disappears. This work suggests that in
compact packed {\it sc} Na$_{2}$AC$_{60}$ \ systems the alkali ions can not
be regarded as lattice spacers only, and they have a direct effect on the
electronic properties. Obviously, theoretical work which considers alkali
specific effects is urgently needed.

(B.R.) and (L.D.) wish to thank J. M\"{u}ller for technical help. G.
Oszl\'{a}nyi is acknowledged for the x-ray characterization of the samples.
Work at EPFL and ETHZ was supported by the Swiss National Foundation for the
Scientific Research. This work is performed within a European TMR Network
FULPROP. The NSF DMR 9803025 is acknowledged for support.

Fig. 1. a.) Spin-susceptibility and b.) ESR\ linewidth of Na$_{2}$CsC$_{60}$(%
$\blacksquare )$. Normalized susceptibility and linewidth of K$_{3}$C$_{60}$
(dashed curve) are shown for comparison. The solid curve is a calculated
susceptibility (see text). (Inset in Fig. 1a. shows the low $T$ behavior of $%
\chi (T)$). Arrows indicate, $T_{c}$, the superconducting transition
temperature and, $T_{s},$ the {\it sc}-{\it fcc} transition transition
temperature.

Fig. 2. a.) Infrared reflectivity and b.) the real part of the optical
conductivity, $\sigma _{1}$($\omega $), of powder (cubic) Na$_{2}$CsC$_{60}$%
. A Lorentz Drude fit at 200 and 20 K\ is also shown (see text). Inset shows
conductivity for the polymerized pellet sample.


\begin{references}
\bibitem{saitocomment}  Samples were always rapidly cooled in Ref.\cite
{maniwaprb95} thus polymeric phases were not present; T. Saito private
communication.
\end{references}
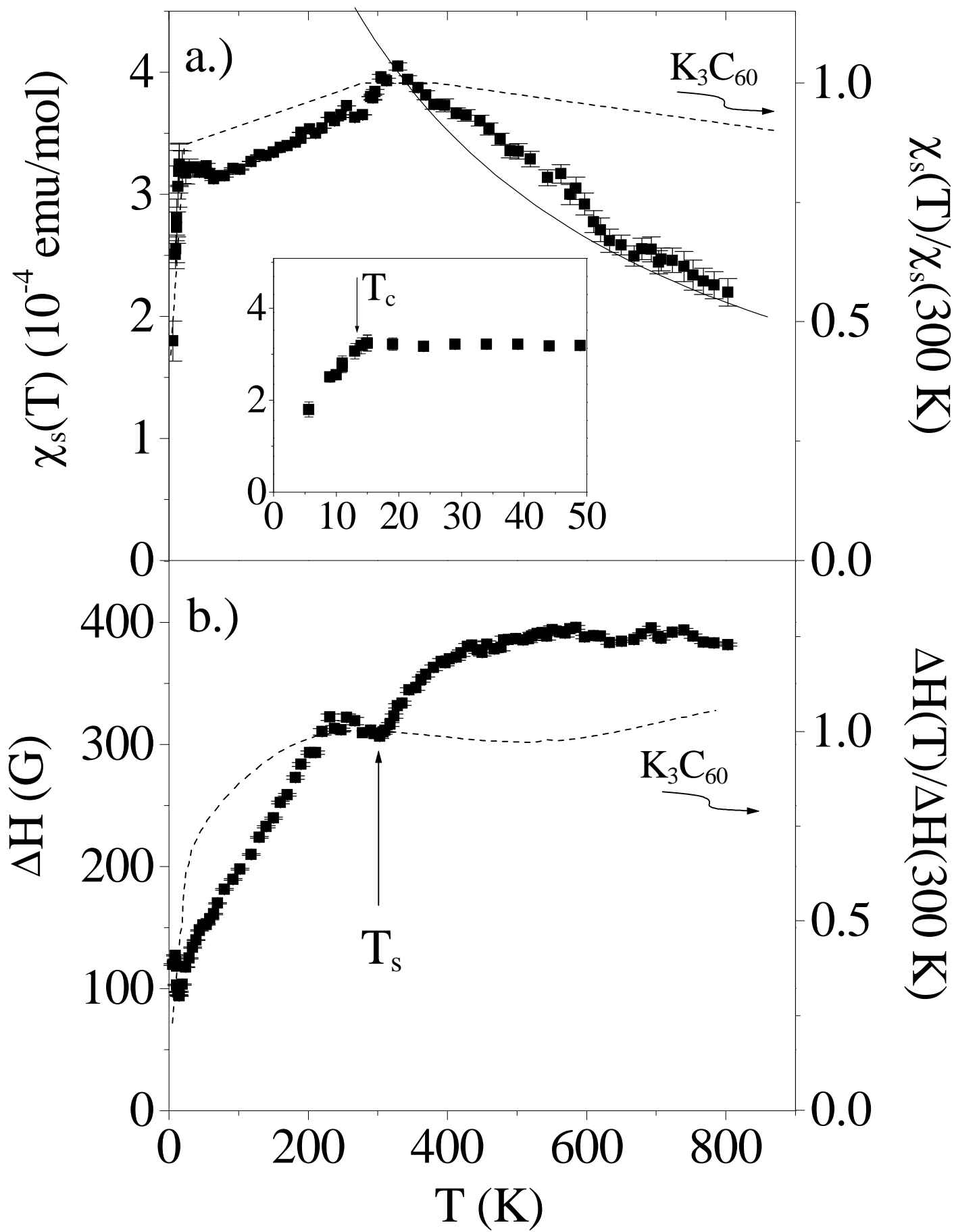
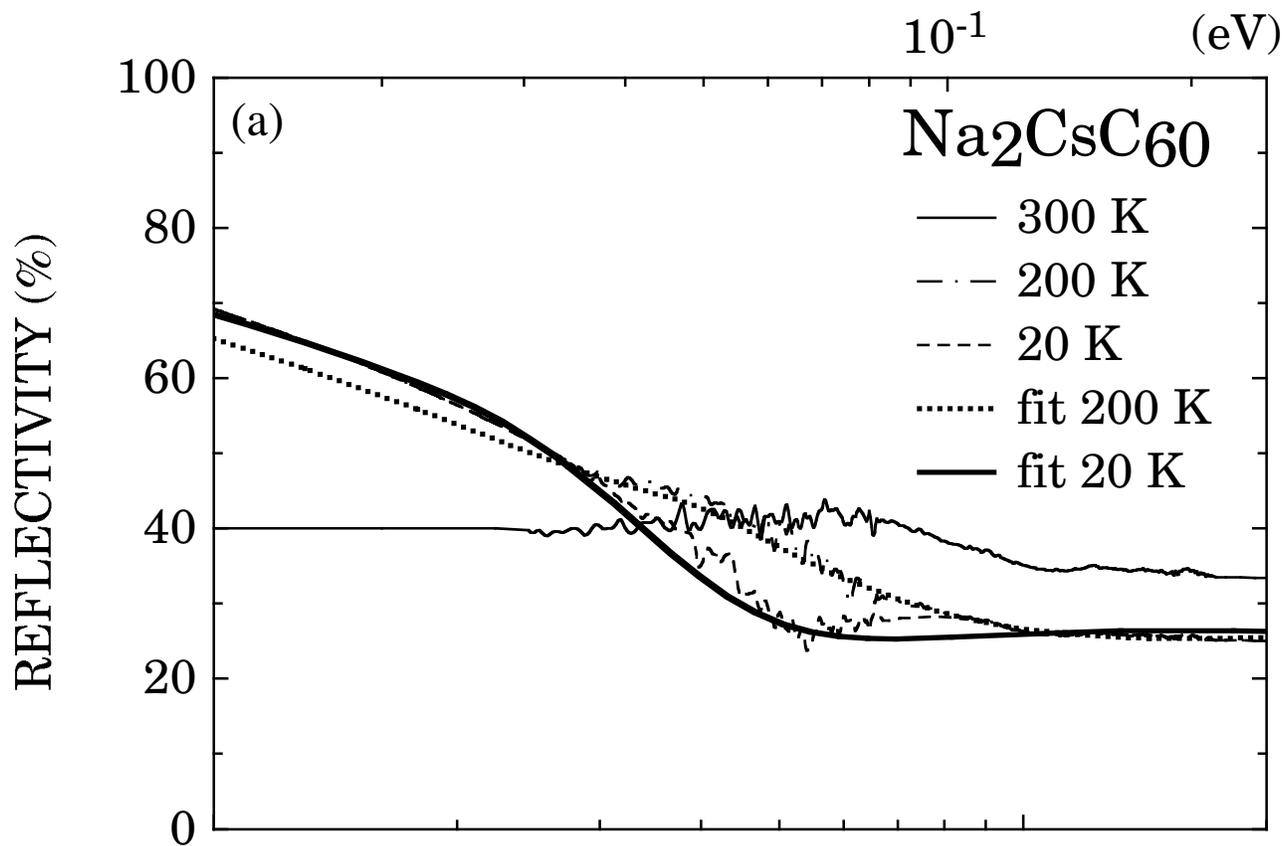
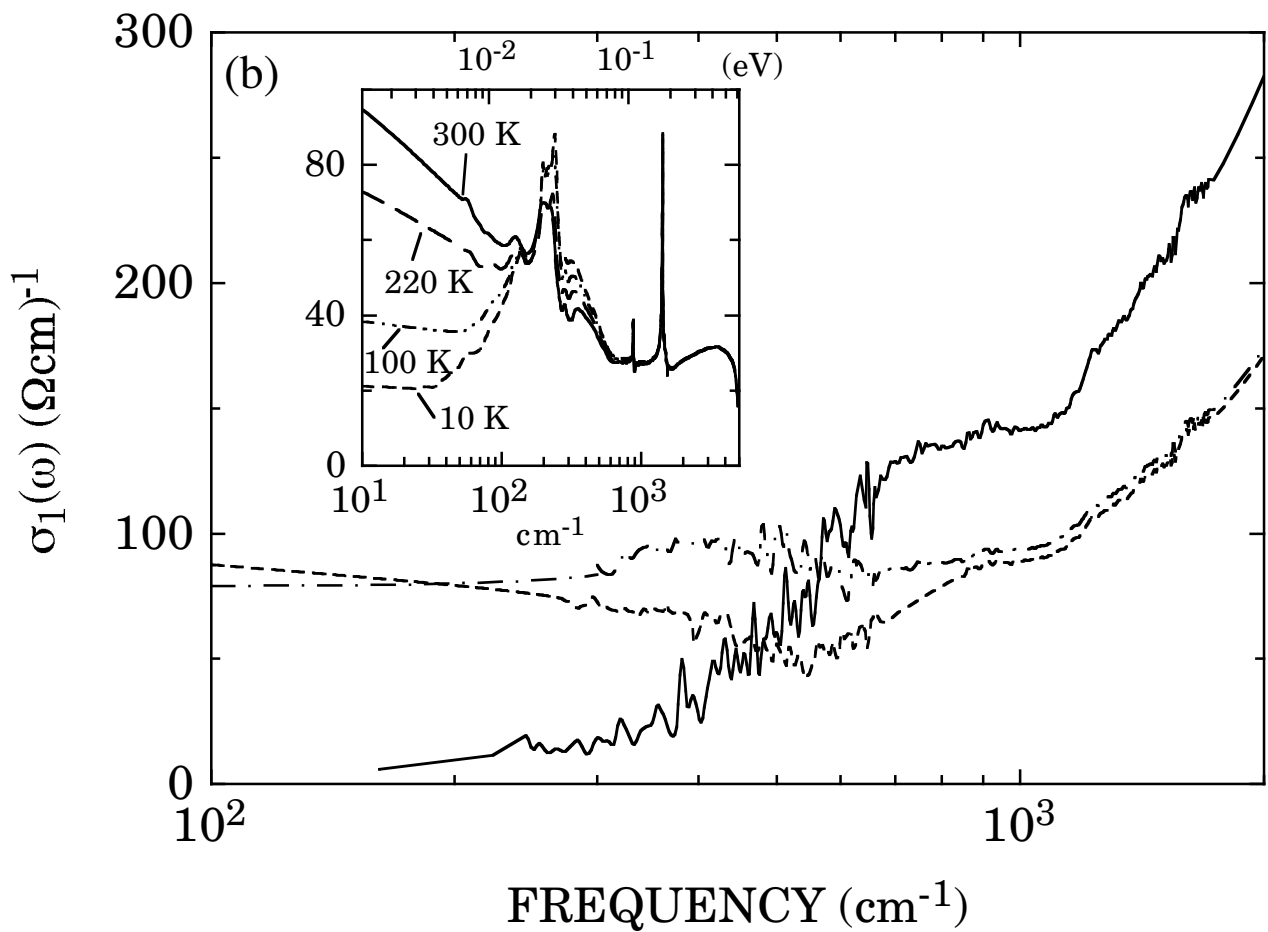
\end{document}